\documentclass[
reprint,
superscriptaddress,
footinbib
amsmath,
amssymb,
aps,
prl,
longbibliography,
]{revtex4-2}
\usepackage{amsfonts}
\usepackage{balance}
\usepackage{graphicx}
\usepackage{epstopdf}
\usepackage{dcolumn}
\usepackage{bm}
\usepackage{hyperref}
\hypersetup{colorlinks=true, linkcolor=cyan, filecolor=cyan, urlcolor=cyan, citecolor=cyan,}
\usepackage[mathlines]{lineno}
\usepackage{mathrsfs}
\usepackage{float}
\usepackage{placeins}
\usepackage{layout}
\usepackage{color}

\begin{document}
	\title{Single Complex-Frequency Resonance Mode in an Engineered Disordered Time-Varying Cavity}
	\author{Bo Zhou}
	\affiliation{State Key Laboratory of Extreme Photonics and Instrumentation, College of Information Science and Electronic Engineering, Zhejiang University, Hangzhou 310027, China}
	\affiliation{International Joint Innovation Center, The Electromagnetics Academy at Zhejiang University, Zhejiang University, Haining 314400, China}
	\author{Xianmin Guo}
	\affiliation{Information Materials and Intelligent Sensing Laboratory of Anhui Province, Anhui University, Hefei 230601, China}
	\author{Xinsong Feng}
	\affiliation{Department of Electrical and Computer Engineering, University of California at Los Angeles, Los Angeles, CA 90095, USA}
	\author{Hongsheng Chen}\email{hansomchen@zju.edu.cn}
	\affiliation{State Key Laboratory of Extreme Photonics and Instrumentation, College of Information Science and Electronic Engineering, Zhejiang University, Hangzhou 310027, China}
	\affiliation{International Joint Innovation Center, The Electromagnetics Academy at Zhejiang University, Zhejiang University, Haining 314400, China}
	\author{Zuojia Wang}\email{zuojiawang@zju.edu.cn}
	\affiliation{State Key Laboratory of Extreme Photonics and Instrumentation, College of Information Science and Electronic Engineering, Zhejiang University, Hangzhou 310027, China}
	\affiliation{International Joint Innovation Center, The Electromagnetics Academy at Zhejiang University, Zhejiang University, Haining 314400, China}
	
	\begin{abstract}
	We propose a straightforward mechanism for achieving unique $k$-space resonance modes in one-dimensional time-varying cavities where periodic temporal modulation creates momentum band gaps through Floquet dynamics. By engineering the synergy between cavity resonance conditions and Floquet mode formation in photonic time crystals, we demonstrate the emergence of a single dominant momentum state that exhibits remarkable robustness against temporal disorder. Through analytical modeling and numerical verification, we show that the interplay between time-varying medium and cavity boundary conditions leads to amplification of specific waves followed by spatial mode selection. This engineered resonance mechanism enables insensitivity to initial wave source configuration and strong temporal disorder immunity. Our findings give a simple mechanism for exploiting narrow momentum bandgaps, and establish a foundation for developing high-quality temporal cavity lasers and advancing extreme temporal predictability in time-modulated systems.
	\end{abstract}
	
	\maketitle
	
	\noindent \textit{Introduction}---Recent advances in photonic time crystals (PTCs) \cite{Reyes_AyonaHalevi_665, PhysRevLett_130_093803, Lustig_18, sciadv_adg7541} and space-time metamaterials \cite{CalozDeck_Léger_537, EmanueleRomain_523, science_aat3100, PhysRevLett_132_263802, ZhuZhao_699, PhysRevApplied_19_064072, MaCui_704} have sparked significant interest in manipulating electromagnetic waves through temporal modulation \cite{Zhou2020, Apffel2022, Ramaccia20181968, Engheta20231190, Rizza2022, Ramaccia20201607, Ramaccia20205836, Castaldi20213687, Vázquez_Lozano2023, Ramaccia2021}. The emerging paradigm of time-varying metamaterials provides unprecedented control over momentum band structures \cite{sciadv_adg7541, GaliffiXu_705, Ren2025, Reyes_AyonaHalevi_665, 7570237, WangGarg_676}, complementing the well-established spatial bandgap engineering in photonic crystals \cite{PhysRevLett_125_133603, d2f4e572e28843d59dabb65757281035, Munzberg_18, Li2020, Cai20215898, Wu2020338, Ozawa2019, Dudley20061135}. Practical implementations of PTCs often face intrinsic limitations: narrow bandwidths imposed by modulation strength constraints \cite{WangGarg_676, Asgari_24, Saha_23} and sensitivity to temporal disorder \cite{ApffelWildeman_647}. While intrinsic resonances \cite{WangGarg_676}, biaxial anisotropy \cite{PhysRevLett_134_063801}, and longitudinal phonon modes \cite{PhysRevB_110_L100306} have been explored to broaden bands and facilitate their utilization, existing approaches remain complex. \par 
	In this Letter, we propose a fundamentally different strategy by establishing a single complex-frequency resonance mode through synergistic cavity-wave dynamics in a simplified 1D time-varying system. Our system consists of a PEC-bounded cavity with instantaneous time-modulated permittivity $\varepsilon(t)$, where periodic temporal interfaces generate Floquet states while cavity boundaries enforce standing wave conditions. The key innovation lies in the self-consistent coupling between wave splitting at multiple temporal interfaces and spatial mode selection through cavity resonance. This dual mechanism produces a unique $k$-space mode that exhibits: independence from initial source configuration due to cavity mode purification, strong resilience against temporal disorder through Floquet state stabilization, and intrinsic momentum space discretization. These characteristics suggest transformative potential for laser cavity design and temporal predictability enhancement in engineered time-varying systems.\par
\begin{figure}[tbp]
		\includegraphics{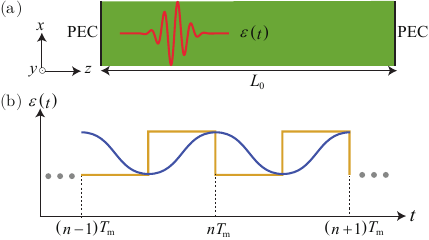}
		\caption{\label{schematic}(a) Schematic of 1D time-varying cavity, the red line is the Gaussian pulse propagating in the cavity and the green zone is the time-varying region; (b) Two different time-varying modes of permittivity, where blue is stepwise and yellow is sinusoidal.}
	\end{figure}

	\noindent\textit{One-dimensional Time-Varying Cavity}---We consider the one-dimensional (1D) time-varying cavity as shown in Fig. \ref{schematic}(a), where ${L_0}$ is the length of the cavity. The ends of the cavity are perfect electric conductors (PEC). Here, for simplicity, the permittivity of the medium filling the cavity is assumed as time-varying [$\varepsilon= \varepsilon_0 \varepsilon(t)$] and the permeability is $\mu=\mu_0$, where $\varepsilon_{0}$ and $\mu_0$ are the permittivity and permeability of the vacuum, respectively. Electric filed is $x$-polarized and denoted by $E_x(z, t)$. We assume that the plane wave has the time variations of the form $\exp(- \mathrm{i} \omega t)$ and the space variations of the form $\exp(\mathrm{i} k_z z)$, where $\omega$ is the angle frequency and $k_z$ is the wave number. The response of the material is instantaneous [$D_x(z, t) = \varepsilon_0 \varepsilon(t) E_x(z, t)$]. It is assumed that a Gaussian pulse propagating in the $+z$ direction is already fully excited at $t = 0$ (the initial conditions are discussed in the Appendix). The permittivity then starts to experience periodic modulations with a period of $T_{\mathrm{m}}$, i.e., $\varepsilon(t + T_{\mathrm{m}}) = \varepsilon(t)$. \par
\begin{figure}[tbp]
		\includegraphics{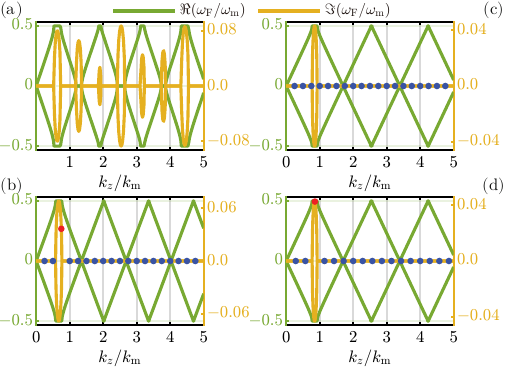}
		\caption{\label{band_structure}(a) Band structure of stepwise modulation; (b) band structure of sinusoidal modulation, where $\varepsilon(t) = \varepsilon_{\mathrm{av}} + \varepsilon_{\mathrm{m}}\cos\left(\omega_{\mathrm{m}} t\right)$, $\varepsilon_{\mathrm{av}} = 2$, $\varepsilon_{\mathrm{m}} = - 1$, $\omega_{\mathrm{m}} = 2 \pi / T_{\mathrm{m}}$ and $k_{\mathrm{m}} = \omega_{\mathrm{m}} \sqrt{\varepsilon_0 \mu_0}$; (c) and (d) band structure of sinusoidal modulation, where $\varepsilon_{\mathrm{av}} = 3$ and $\varepsilon_{\mathrm{m}} = - 1$. Green lines denote the real part and yellow lines denote the imaginary part. In (b) and (c), dots indicate that the corresponding $k_z$ satisfies Eq.~(\ref{standing_condition_of_kz}) when $l = 2$; in (d), dots indicate that the corresponding $k_z$ satisfies Eq.~(\ref{standing_condition_of_kz}) when $l = 1.75$.}
	\end{figure}

	Since the cavity is PEC at both ends, we can consider the cavity as a 1D PTC \cite{PhysRevA_Bozhou}. It is well known that, according to Bolch-Floquet theorem, a periodic change in space opens energy band gaps \cite{RevModPhys_91_015006} and a periodic change in time opens momentum band gaps \cite{Lustig_18, Asgari_24}. In the PTC, according to Floquet theorem \cite{Asgari_24}, we can express $E_x(z, t)$ as 
	\begin{equation}
		E_x(z, t) = E_{\mathrm{F}}(t) \exp(\mathrm{i} \omega_{\mathrm{F}} t) \exp(\mathrm{i} k_z z) 
		\label{Floquet_expression_of_Ex}
		,
	\end{equation}
	where $ E_{\mathrm{F}}(t)$ is  periodic in time with period $T_{\mathrm{m}}$ and $\omega_\mathrm{F}$ is the so-called Floquet frequency. Combining Eq.~(\ref{Floquet_expression_of_Ex}) and Maxwell's equations, we will obtain a eigenvalue equation with respect to $k_z$ and $\omega_\mathrm{F}$ (see Appendix B for details).  Solving this eigenvalue equation, we can obtain the relationship between $k_z$ and $\omega_\mathrm{F}$, which is the so-called band structure. For example, when the $\varepsilon$ experience stepwise modulation, the obtained band structure is shown in Fig. \ref{band_structure}(a); when the $\varepsilon$ experience sinusoidal modulation, the obtained band structure is shown in Fig. \ref{band_structure}(b) and (c). We tend to be concerned with momentum bandgaps where $\Im [\omega_\mathrm{F}(k_z)]$ is not zero. Two different patterns of fields emerge from these bandgaps: one exponentially grow ($\Im [\omega_\mathrm{F}(k_z)] < 0$) and another exponentially decay ($\Im [\omega_\mathrm{F}(k_z)] > 0$) over time, due to the energy non-conservation in PTC. \par
	\noindent\textit{Single Resonance Mode: Standing Wave}---Typically, the momentum bandgap is desired to be wide enough and numerous enough so as to facilitate our manipulation of the wave in the bandgap \cite{WangGarg_676}. This requires large modulation depths and modulation frequencies with high order harmonics. Stepwise modulation contains an infinite number of harmonics, and thus it has an infinite number of momentum bandgaps [Fig.~\ref{band_structure}(a)]; single-harmonic modulation, accordingly, will only open a unique momentum bandgap [Fig.~\ref{band_structure}(b)]. The former is much more difficult to realize experimentally than the latter. Here, we propose a mechanism which utilizes cavity resonance combined with a unique momentum bandgap for single harmonic modulation [$\varepsilon(t) = \varepsilon_{\mathrm{av}} + \varepsilon_{\mathrm{m}}\cos\left(\omega_{\mathrm{m}} t\right)$] to manipulate wave modes on $k$-space. It is well known that waves in time-varying media will experience time interfaces and split into forward and backward waves \cite{Asgari_24, Lustig_18, 2022_AP, JonesKildishev_706, GaliffiXu_705, PhysRevLett_132_263802, MoussaXu_701, Pacheco_PeñaKiasat_677}. The multiple time interfaces brought about by the periodic time-varying media, which repeatedly act on the split forward and backward waves, lead to an exponential increase in the number of waves \cite{PhysRevA_Bozhou}. The final superposition of these waves in the cavity will eventually result in a standing wave pattern in space due to resonance. The standing wave pattern requires that $\lambda_z = \frac{2 \pi}{k_z} = \frac{2 L_0}{q}$, $q \in \mathbb{N^+}$, where $\lambda_z$ is the wavelength corresponding to $k_z$. Let $L_0 = l \lambda_{\mathrm{m}}$ ($l$ is an arbitrary constant greater than $0$, and $\lambda_{\mathrm{m}} = \frac{2 \pi}{k_{\mathrm{m}}}$ is  the wavelength corresponding to $k_{\mathrm{m}}$), then
	\begin{equation}
		k_z = \frac{2 q \pi}{2 L_0} = \frac{q k_{\mathrm{m}}}{2 l}, ~q \in \mathbb{N^+}.
		\label{standing_condition_of_kz}
	\end{equation}
	We do not make any other special design for the PTC. Then the decaying and growing modes will be excited simultaneously in the momentum bandgap. Since the decay rate for the decaying mode and the growth rate for the growing mode are both exponential over time. Thus we will end up observing that the final standing wave modes formed in the cavity are those whose $k_z$ satisfies Eq.~(\ref{standing_condition_of_kz}). Other modes are completely suppressed. \par
	As previously stated, single-harmonic sinusoidal modulation will only open a single momentum bandgap, and this bandgap is finite in width. Therefore, we can adjust the size of the cavity length $L_0$ (i.e., $l$) in the Eq.~(\ref{standing_condition_of_kz}) so that the $k_z$ of the growing mode wave that satisfies the resonance condition is unique. For example, we let $\varepsilon_{\mathrm{av}} = 2$, $\varepsilon_{\mathrm{m}} = - 1$ and $l = 2$. Then we obtain only one $k_z = 0.75 k_\mathrm{m}$ which falls in the unique momentum band gap, marked with a red dot in Fig.~\ref{band_structure}(b). This means that the initial Gaussian pulse will  eventually evolve into spatial standing waves with $k_z = 0.75 k_\mathrm{m}$.
	\par
	In order to observe the proposed theoretical predictions, three different modified Finite-Difference Time-Domain (FDTD) methods and the time-domain solver of COMSOL Multiphysics are used for simulations (see Appendix A for details of the simulation setup). The simulation results show that all four different computational methods yield consistent phenomena, so the subsequent presentation of the results will only show the results of one simulation method, and the rest of the results are shown in the Appendix. \par
	As shown in Fig.~\ref{result_1}(a) and (b), the initial Gaussian pulse evolves into a fixed spatial standing wave with a predicted wavelength ${\lambda _z} = \frac{{2\pi }}{{{k_z}}} = \frac{4}{3} \lambda _{\text{m}} \approx 1.33\lambda _{\text{m}}$ after about $20 T_\mathrm{m}$.
	We define the characteristic multipliers $\rho$ as $\rho = \exp(\mathrm{i} \omega_{\mathrm{F}} T_\mathrm{m}) = \left|\rho\right| \exp(\mathrm{i} \theta)$ to measure the evolution of $\left|E_x\right|$ with time, then it is easy to obtain that $\left|\rho\right| = \left|\max\limits_z[E_x(z, t + T_{\rm m})]/\max\limits_z[E_x(z, t)]\right|$ in the simulation. Then we can find that the $\left|\rho\right|$ obtained from the simulation [blue line in Fig.~\ref{result_1}(c)] and the $\left|\rho\right|$ obtained from the theoretical calculation [red dashed line in Fig.~\ref{result_1}(c)] are in good agreement, i.e., the gain obtained from the time-varying medium for the standing wave is in accordance with the prediction obtained from Floquet theorem [Eq.~(\ref{Floquet_expression_of_Ex})]. Also we have $\theta = \Re(\omega_{\mathrm{F}}) T_\mathrm{m} = 0.5 \omega_{\mathrm{m}} T_\mathrm{m} = \pi$, which is why in Fig.~\ref{result_1}(a) the electric field always changes sign every $T_\mathrm{m}$ (i.e., it alternates between yellow and blue). We can also demonstrate this phenomenon with the 1D electromagnetic energy. Here, the 1D electromagnetic energy \footnote{Strictly speaking, this would be the surface electromagnetic energy density in the $x\mathrm{-}y$ plane. However, we are considering a 1D cavity, so this does not prevent us from using this definition to discuss 1D energy.} is defined as $U(t) = \left[2 \varepsilon _0 \varepsilon (t)\right]^{ - 1} \int {\left| D_x(z, t) \right|}^2 \mathrm{d}z + \left( 2 \mu _0 \right)^{ - 1} \int {\left| B_y(z,t) \right|}^2 \mathrm{d}z$ \cite{SharabiLustig_645}. With the help of  Eq.~(\ref{Floquet_expression_of_Ex}), we can find that for the electromagnetic fields at the momentum bandgap, $\ln[U(t)] \propto t$ \cite{PhysRevA_Bozhou, SharabiLustig_645, Lustig_18}. In Fig.~\ref{result_1}(e), our simulation results also show the same phenomenon that standing waves within the momentum bandgap can stably extract energy from time-varying media, with the energy increasing exponentially over time. \par
	\begin{figure*}[tbp]
		\includegraphics{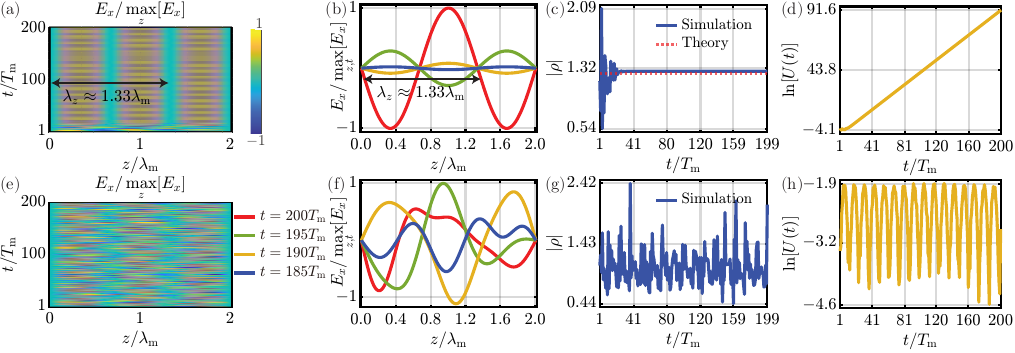}
		\caption{\label{result_1} When $\varepsilon_{\mathrm{av}} = 2$, $\varepsilon_{\mathrm{m}} = - 1$, and $l = 2$, (a) temporal evolution of the space distribution of the normalized electric field, (b) space distribution of the normalized electric field at several selected specific times, (c) characteristic multipliers obtained from theoretical calculations and simulations, and (d) temporal evolution of the 1D EM energy. When $\varepsilon_{\mathrm{av}} = 3$, $\varepsilon_{\mathrm{m}} = - 1$, and $l = 2$, (e) temporal evolution of the space distribution of the normalized electric field, (f) space distribution of the normalized electric field at several selected specific times, (g) characteristic multipliers obtained from the simulations, and (h) temporal evolution of the 1D EM energy. In all of these subfigures, $E_x$ implicitly expresses $E_x(z, t)$. An enlarged version of (a) can be found in the Appendix.}
	\end{figure*}
	However, when we change $\varepsilon_{\mathrm{av}}$ to $\varepsilon_{\mathrm{av}} = 3$, the momentum bandgap will not contain any $k_z$ that satisfies Eq.~(\ref{standing_condition_of_kz}), as shown in Fig.~\ref{band_structure}(c). That is, ultimately no resonant modes (standing wave) are available to derive gain from the time-varying medium. As shown in Fig.~\ref{result_1}(e) and (f), even after $200 T_\mathrm{m}$, the same initial Gaussian pulse still fails to evolve to one of the or several standing waves. Meanwhile, comparing Fig.~\ref{result_1}(c) and Fig.~\ref{result_1}(g), Fig.~\ref{result_1}(g) is almost always repeating the unstable situation at the first $20 T_\mathrm{m}$ of Fig.~\ref{result_1}(c). This indicates that the Gaussian pulse is unable to achieve stable gain from the time-varying medium when there is no $k_z$ in the momentum bandgap that satisfies Eq.~(\ref{standing_condition_of_kz}). As can also be observed in Fig.~\ref{result_1}(h), the electromagnetic energy is in a oscillation state. Therefore, there is no stable Floquet mode in this cavity, even if the Gaussian pulse contains components of $k_z$ that fall in the momentum band gap. \par
	Now, we have shown that in the ordered time-varying cavity, the resonant modes can be manipulated on $k$-space using any single momentum bandgap as long as the cavity length $L_0$ is rationally designed via Eq.~(\ref{standing_condition_of_kz}). We can choose whether or not a resonant mode exists as well as ensure that the resonant mode is unique. However, it is imperative to acknowledge that real-world systems are invariably subject to disorder. This inherent disorder can significantly influence the dynamics of systems and potentially obscure or modify the observed effects. Therefore, to bridge the gap between theoretical insights and practical applications, it is crucial to investigate how the identified phenomena manifest in the presence of disorder. This exploration will not only enhance our understanding of the robustness of these effects but also provide valuable guidance for their exploitation in realistic, disordered time-varying cavity systems. Consequently, the next step in our analysis involves examining the interplay between disorder and the time-varying properties of the cavity, aiming to uncover strategies to harness these phenomena in practical scenarios. \par
	\noindent\textit{Disordered Time-Varying Cavity}---To study the effect of disorder on the resonant modes, we consider the following time-varying $\varepsilon(t)$:
	 \begin{equation}
	 	\varepsilon(t) = \varepsilon_{\mathrm{av}} + \varepsilon_{\mathrm{m}}\cos\left(\omega_{\mathrm{m}} t\right) + \varepsilon_{\mathrm{d}} R(t)
	 	\label{eps1}
	 	.
	 \end{equation}
	For simplicity, we assume that $R(t)$ is a uniformly distributed stationary random process on $[-1, 1]$. $R(t)$ is designed to have only short-range autocorrelation, preventing the introduction of extreme high-frequency components [see Appendix C for the discussion of  $R(t)$]. We let $\varepsilon_{\mathrm{av}} = 3$, $\varepsilon_{\mathrm{m}} = - 1$ and $l = 1.75$. When there is no disorder (i.e., $\varepsilon_d = 0$), then we obtain only one $k_z = \frac{12}{14} k_\mathrm{m} \approx 0.857 k_\mathrm{m}$ which falls in the unique momentum band gap, marked with a red dot in Fig.~\ref{band_structure}(d). That is, the initial Gaussian pulse will evolve into a fixed spatial standing wave with a predicted wavelength ${\lambda _z} = \frac{{2\pi }}{{{k_z}}} = \frac{14}{12} \lambda _{\text{m}} \approx 1.17\lambda _{\text{m}}$. To show that our proposed mechanism is strongly resistant to disorder, we let $\varepsilon_d = \varepsilon_m = - 1$, i.e., strongly disordered. \par
\begin{figure*}[tbp]
		\includegraphics{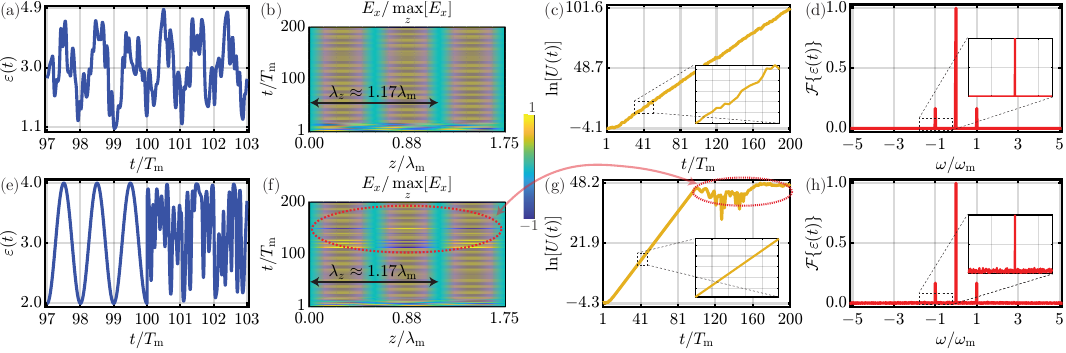}
		\caption{\label{result_2} When the the time-varying $\varepsilon(t)$ follows Eq.~(\ref{eps1}): (a) one sample of Eq.~(\ref{eps1}), (b) temporal evolution of the spatial distribution of the normalized electric field, and (c) temporal evolution of the 1D EM energy. (d) $\mathcal{F}\{\varepsilon(t)\}$ when $\varepsilon_d = 0$. When the the time-varying $\varepsilon(t)$ follows Eq.~(\ref{eps2}): (e) one sample of Eq.~(\ref{eps2}), (f) temporal evolution of the spatial distribution of the normalized electric field, and (g) temporal evolution of the 1D EM energy. (h) $\mathcal{F}\{\varepsilon(t)\}$ when $\varepsilon_d = - 1$. An enlarged version of (f) can be found in the Appendix.}
	\end{figure*}

	One sample of $\varepsilon(t)$ is presented in Fig.~\ref{result_2}(a), with the corresponding temporal evolution of the spatial distribution of the electric field and the 1D EM energy shown in Figs.~\ref{result_2}(b) and (c), respectively. It is evident that even under strong disorder perturbations, the initial Gaussian pulse evolves into the designed resonant mode—a spatial standing wave with ${\lambda_z} \approx 1.17\lambda_{\text{m}}$. From the perspective of 1D electromagnetic energy evolution, disorder introduces slight deviations to the Floquet mode, transforming the originally strict linear relationship [Fig.~\ref{result_1}(d)] into an approximately linear trend [Fig.~\ref{result_2}(c)]. We attribute the robustness of this single resonant mode in the time-varying cavity against disorder to two primary reasons. The first reason is that disorder itself does not significantly disrupt periodicity. This can be explained from the frequency-domain perspective. Fig.~\ref{result_2}(d) shows the Fourier transform $\mathcal{F}\{\varepsilon(t)\}$ when $\varepsilon_d = 0$, while Fig.~\ref{result_2}(h) shows $\mathcal{F}\{\varepsilon(t)\}$ for $\varepsilon_d = -1$. Notably, the frequency components of $R(t)$ are broadly distributed across a wide range, exerting minimal influence on the dominant frequency components of $\varepsilon(t)$ ($\omega = 0$ and $\omega = \pm \omega_{\text{m}}$). Since the formation of Floquet modes primarily relies on the periodicity of the $\omega = \pm \omega_{\text{m}}$ components and cavity resonance modes are discrete and invariant, the Floquet mode persists and ultimately stabilizes into a single resonant mode.\par 
	The second reason is the inherent immunity of the resonant mode—spatial standing waves—to temporal disorder. For a spatial standing wave $E_x$ in a time-varying cavity, it can always be expressed as $E_x(k_z) = \varphi(k_z) + \varphi(-k_z)$, where $\varphi(k_z)$ represents the component propagating along $+z$, and $\varphi(-k_z)$ corresponds to the $-z$ component. The time-varying nature of the medium causes $\varphi(k_z)$ to branch into $\mathcal{T}\varphi(k_z)$ and $\mathcal{R}\varphi(-k_z)$, while $\varphi(-k_z)$ splits into $\mathcal{T}\varphi(-k_z)$ and $\mathcal{R}\varphi(k_z)$, where $\mathcal{T}$ and $\mathcal{R}$ are the transmission and reflection coefficients due to time-varying \cite{Asgari_24, Lustig_18, 2022_AP, JonesKildishev_706, GaliffiXu_705, PhysRevLett_132_263802, MoussaXu_701, Pacheco_PeñaKiasat_677, PhysRevA_Bozhou}. Since these splittings occur simultaneously, the resultant field $E_x(k_z) = \mathcal{T}\varphi(k_z) + \mathcal{R}\varphi(-k_z) + \mathcal{T}\varphi(-k_z) + \mathcal{R}\varphi(k_z) = \mathcal{T}[\varphi(k_z) + \varphi(-k_z)] + \mathcal{R}[\varphi(-k_z) + \varphi(k_z)]$ remains a standing wave. To validate this, we employ the following configuration:  
	\begin{eqnarray}
		\varepsilon(t) &=& \varepsilon_{\mathrm{av}} + \varepsilon_{\mathrm{m}} \cos\left(\omega_{\mathrm{m}} t\right) \left[f_{H}(t) - f_{H}(t - \kappa T_{m})\right] \nonumber\\
		&& + \varepsilon_{\mathrm{d}} R(t) f_{H}( t - \kappa T_{m})
		\label{eps2}
		,
	\end{eqnarray}
	where $f_{H}(\cdot)$ denotes the Heaviside step function. Setting $\kappa = 100$, $\varepsilon$ undergoes ordered modulation when $t < 100T_{\mathrm{m}}$ to establish a stable resonant mode, followed by fully disordered temporal variations when $t > 100T_{\mathrm{m}}$. A sample of $\varepsilon(t)$ for $\kappa = 100$ is shown in Fig.~\ref{result_2}(e), with the corresponding temporal evolution of the spatial distribution of the electric field and the 1D EM energy illustrated in Figs.~\ref{result_2}(f) and (g). During the initial $100T_{\mathrm{m}}$, the spatial standing wave remains stable due to Floquet mode formation, and $\ln[U(t)]$ exhibits strict linear growth. Beyond $100T_{\mathrm{m}}$, the system becomes fully disordered, leading to the disappearance of the Floquet mode and the loss of linear growth of $\ln[U(t)]$. However, the standing wave pattern remains stable. Close inspection of the red-circled region in Fig.~\ref{result_2}(f) reveals that the electric field no longer alternates polarity every $T_m$ interval, confirming the absence of the Floquet mode. This verifies the temporal disorder immunity of resonant standing waves in time-varying cavities. \par
	For the phenomenon described in this letter, it appears that the fluctuations of $\varepsilon$ must be fast (on the scale of the electric period) if we want to realize it in practice. This requires a discussion of possible experimental realization schemes. Since we are considering a 1D cavity,  a experimental schemes based on the naturally 1D structures--transmission line metamaterial \cite{Reyes_AyonaHalevi_665, HuidobroGaliffi_697, MoussaXu_701, PhysRevLett_121_204301, GaliffiXu_705, JonesKildishev_706} has been discussed in the Appendix. \par
	\noindent\textit{Summary}---We have demonstrated a fundamental mechanism for generating single complex-frequency resonance modes in temporally modulated electromagnetic cavities with and without disorder. The proposed 1D cavity with time-varying permittivity achieves unique $k$-space mode selection through the self-organization of Floquet states and cavity resonance conditions. This mode selection is extremely resistant to temporal disorder. This discovery addresses two critical challenges in time-modulated photonics: overcoming narrow Floquet band limitations and mitigating temporal disorder effects. The demonstrated principles provide a roadmap for developing (i) ultra-stable time-crystal lasers with intrinsic mode selection and (ii) precision temporal cavities for fundamental tests of time-domain quantum electrodynamics. The 1D implementation scheme using transmission line metamaterial suggests immediate experimental feasibility with existing modulation technologies.\par
	The work was sponsored by the National Natural Science Foundation of China (62222115, 62171407), the Key Research and Development Program of Zhejiang Province under Grant No.2024C01241(SD2), and the Fundamental Research Funds for the Central Universities.\par

%

\end{document}